# Time series and the meaning of quantum non-locality.

Alejandro A. Hnilo.

CEILAP, Centro de Investigaciones en Láseres y Aplicaciones, (CITEDEF-CONICET).
CITEDEF, J.B. de La Salle 4397, (1603) Villa Martelli, Argentina.
*email*: ahnilo@citedef.gob.ar

*Abstract*

Quantum non-locality has become a popular term. Yet, its precise meaning, and even its mere existence, is the subject of controversies. The main cause of the controversies is the never ending discussion on the appropriate definitions of Locality and Realism in the derivation of Bell's inequalities. On the other hand, Louis Sica derived Bell's inequalities from the hypothesis that time series of outcomes observed in one station do not change if the setting in the other (distant) station is changed. The derivation is based on arithmetical properties of time series only; it does not involve the definitions of Locality, Realism and probabilities, and is valid for series of any length. An important consequence is that violation of Bell's inequalities implies the series recorded (at the same time) in one station to be different if the setting in the other station is changed. This result gives precise meaning to the idea of quantum non-locality, and also makes evident why using it for faster than light signaling is impossible (because the difference is between factual and counter-factual series). Finally, it is demonstrated that series of outcomes, even if they violate Bell's inequalities, can be always embedded in a set of factual and counter-factual data that does hold to Bell's inequalities. Loosely speaking, the factual world may be quantum or not, but the union of factual and counter-factual worlds is always classical.

*October 18th, 2024.*

**Introduction.**

The experimental violation of Bell's inequalities means that at least one of the hypotheses involved in their derivation is false in Nature. The best known hypotheses can be gathered in two large groups: "Locality" and "Realism". Therefore, one of the interpretations of the violation of Bell's inequalities implies the existence of non-local effects (i.e., effects that propagate at infinite speed), in clear conflict with the Special Theory of Relativity. The conflict is usually circumvented by noting that these effects, according to standard Quantum Mechanics, do not allow transmitting information ("no-signaling"). Yet, the tension between the two theories remains. That's why E.T. Jaynes famously stated that Quantum Mechanics and Relativity are in "peaceful coexistence", a term which was coined to describe the fragile relationship between the East and West political and military blocks during the Cold War.

The conflict is not fully solved, among other reasons, because there is no universal agreement on the precise definitions of Locality and Realism. The meaning and consequences of these hypotheses have been the subjects of discussions for years. In particular, many researchers (including myself) have consistently argued that quantum non-locality is an illusion produced by the wave (in general: non-Boolean) nature of quantum systems, and that it does not really exist [1-5]. Of course, this argument depends on what is exactly understood by "non-locality".

Nevertheless, there is a derivation of Bell's inequalities that is free from philosophical and mathematical intricacies, from the definition of classical probabilities and from the size of the statistical sample. This derivation is due to Louis Sica [6], and is based only on arithmetical properties and a well defined intuitive condition. It does not deal with average rates as most usual derivations, but with *time series* of observed outcomes. These series are indisputable experiments' output, so that the conclusions reached by this approach are hard to object.

In this paper, it is shown that Sica's approach assigns "quantum non locality" a clear and incontestable meaning, and explains why this non-locality is useless to send signals. In the next Section 1, the basics of time series in time-stamped Bell's experiments, and *Sica's condition*, are reviewed. In Section 2, the reason why Bell's inequalities can be violated is explained; it is concluded that giving up the possibility of arbitrary reordering the time series plays the role of giving up Local Realism. Section 3 deals with the meaning of quantum non-locality and the (im)possibility of using it to send signals. Section 4 shows that experimentally recorded series may belong to the classical world, or not (that is, they may hold to Bell's inequalities, or not), but that they always belong to a factual + counterfactual world that is classical (that is, the recorded series are always part of extended series that do hold to Bell's inequalities). In other words: there is always a table of (factual + counterfactual) data that *explains*, in classical terms, the violation of Bell's inequalities.

## 1. Time series produced in a (time stamped) Bell's experiment.

Let consider the idealized Bell's (or Einstein-Podolsky-Rosen-Bohm) optical time-stamped experiment in Figure 1. The source S emits two beams entangled in polarization towards stations Alice (A) and Bob (B). In each station, polarization analyzers set at angles $\{\alpha,\beta\}$ split the beams to single photon detectors named "+" and "-". Each photon's *time value* of detection is saved in appropriate devices. The output of this setup are hence time series of outcomes, rather than numbers of detections [7,8]. In Fig.1, the time series $a_i$ of outcomes in station A is (-1, +1, 0, -1, +1, +1); the time series $b_i$ in station B is (-1, -1, +1, 0, +1, 0). Coincidences occur when $a_i.b_i \neq 0$, in this case, in the time slots $i=\{0,1,4\}$. After the experimental run has ended, numerical processing of the recorded time series allows computing Bell's inequalities.

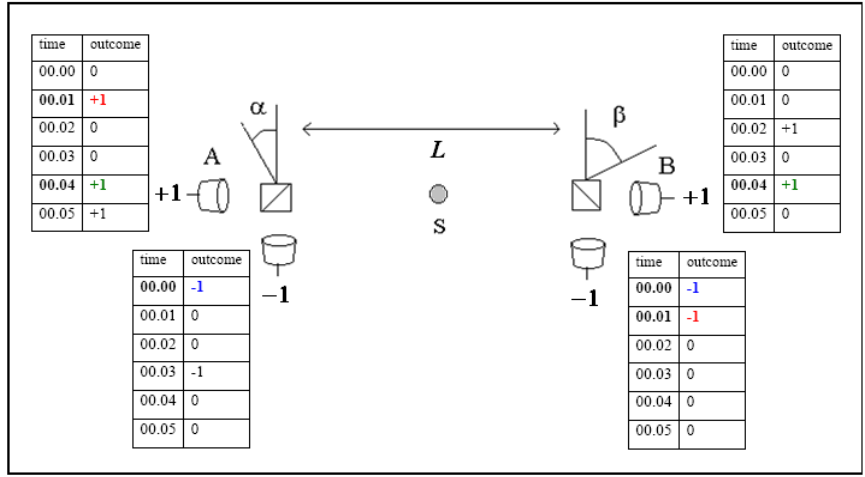

Figure 1: Sketch of a time-stamped Bell's experiment. The source S emits entangled beams that propagate towards stations A and B, which are separated by a (large) distance *L*. The "+1" (-1) in the tables means that detector "+1" (-1) fired during that time slot, "0" means that no detector fired. In this illustration, a coincidence (-,-) occurs at t=00.00, a (+,-) at t=00.01, a (+,+) at t=00.04. Single detections occur at t=00.02 (0,+), t=00.03 (-,0), etc. The rate coincidences/singles defines the efficiency η of each detector. F.ex., the time series of detector A+ has 3 singles and 2 coincidences, then $\eta^+_A = 2/3$.

Computing Bell's inequalities requires recording series for different angle settings. But it is impossible recording with different angle settings (say, A=α and A=α') during the same time slot. A typical assignment of time during an experimental run of total duration T is shown in the upper part of Figure 2. A possible table of actually recorded series of outcomes is shown in the lower part. The $S_{CHSH}$ parameter of the Clauser-Horne-Shimony-Holt form of Bell's inequalities (CHSH) [9] is defined as:

$$(1/N).|\Sigma\, a_i.b_i - \Sigma\, a_i.b'_i| + (1/N).|\Sigma\, a'_i.b_i + \Sigma\, a'_i.b'_i| \equiv$$
$$= |E(\alpha,\beta) - E(\alpha,\beta')| + |E(\alpha',\beta) + E(\alpha',\beta')| \equiv S_{CHSH} \quad (1)$$

The empty boxes in Fig.2 correspond to non-performed observations. F.ex: for time slots from $i$=9 to 16, the setting is A=α', so that there are *no outcomes* (neither +,-, or 0) for the elements $a_9$ to $a_{16}$. For this reason, strictly speaking, $S_{CHSH}$ cannot be calculated from the data in Fig.2 (see Section 4). If the calculation is anyway made as usual (series' length $N$=4), then $S_{CHSH}$ = 2 in this case.

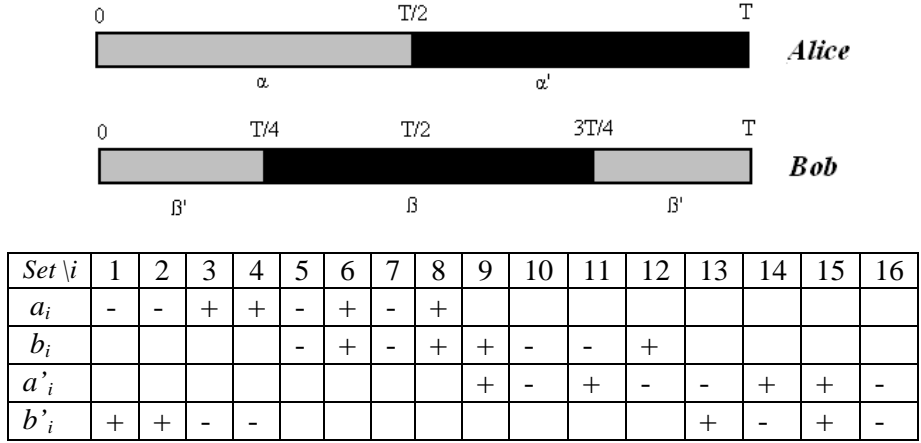

| Set \i | 1 | 2 | 3 | 4 | 5 | 6 | 7 | 8 | 9 | 10 | 11 | 12 | 13 | 14 | 15 | 16 |
|---|---|---|---|---|---|---|---|---|---|---|---|---|---|---|---|---|
| $a_i$ | - | - | + | + | - | + | - | + |  |  |  |  |  |  |  |  |
| $b_i$ |  |  |  |  | - | + | - | + | + | - | - | + |  |  |  |  |
| $a'_i$ |  |  |  |  |  |  |  |  | + | - | + | - | - | + | + | - |
| $b'_i$ | + | + | - | - |  |  |  |  |  |  |  |  | + | - | + | - |

Figure 2: Up: Typical distribution of total recording time T among the different angle settings necessary to test Bell's inequalities. F.ex: A is set equal to α between t=0 and t=T/2 (gray area in line "Alice"), B is set equal to β between t = T/4 and t = 3T/4 (black area in line "Bob"). Down: a possible series of outcomes, if recording time is assigned as in the upper part of this Figure. Note there are no "0" in the series, these data have ideal efficiency η=1.The empty boxes are not "0", they correspond to non-performed observations. They can be filled only after a "possible world" is assumed [10], meaning that information external to the experiment must be provided, see Section 4.

Let introduce the following intuitive assumption, that is called here *Sica's condition*:

- *The time series $a_i$, $a'_i$ ($b_i$, $b'_i$) are the same regardless* B = β *or* β' (A = α *or* α').

However, long series recorded at different times have a vanishingly small probability of being identical, so that this condition must be somehow relaxed. F.ex, in Fig.2, when B=β' $a_i$ is (-,-,+,+) but it is (-,+,-,+) when B=β'. If Sica's condition is true the series recorded at different times may be scrambled, but they are essentially the same (f.ex: a shuffled set of playing cards is the same as it was before unpacked). In consequence, it must be possible *reordering* the series to make them identical. "Reordering" here means that the sub-indexes $i$ of the series (f.ex.) $a_i$ recorded between $T$/4 and $T$/2 (i.e, when B=β) are changed so that it becomes equal, element by element, to the series $a_i$ recorded between 0 and $T$/4 (i.e, when B=β'). For long and more or less balanced recorded series, this reordering seems always possible. In the worst case, a statistically irrelevant number of elements should be discarded [6]. Yet, there is an important *restriction*: correlations between measured outcomes cannot change, for they are physically real. In consequence, if the series in one station (say, $a_i$) is reordered, the series simultaneously recorded in the other station ($b_i$) must be reordered in the same way, so that the product $a_i b_i$ for the new value of $i$ is the same as for the old value of $i$.

After successful reordering, the table has redundant information, for all series appear twice. The same information can be then stored in a smaller table. F.ex, after reordering the series in Fig.2 so that $a_i$ when B=β' becomes equal to $a_i$ when B=β: (-,+,-,+) and $b_i$ when A=α' becomes equal to $b_i$ when A=α: (-,+-,+), all series appear twice, see Figure 3 (up). The table can be then *condensed* to Figure 3 (down). Note the series' lengths halved (from 8 to 4) and the empty boxes disappeared.

| Set \ $i$ | 1 | 2 | 3 | 4 | 5 | 6 | 7 | 8 | 9 | 10 | 11 | 12 | 13 | 14 | 15 | 16 |
|---|---|---|---|---|---|---|---|---|---|---|---|---|---|---|---|---|
| $a_i$ | - | + | - | + | - | + | - | + | | | | | | | | |
| $b_i$ | | | | | - | + | - | + | - | + | - | + | | | | |
| $a'_i$ | | | | | | | | | - | + | + | - | - | + | + | - |
| $b'_i$ | + | - | + | - | | | | | | | | | + | - | + | - |

| Set\$i$ | 1 | 2 | 3 | 4 |
|---|---|---|---|---|
| $a_i$ | - | + | - | + |
| $b_i$ | - | + | - | + |
| $a'_i$ | - | + | + | - |
| $b'_i$ | + | - | + | - |

Figure 3: If Sica's condition is valid, then the series in Fig.2 can be reordered; the new table has redundant lines (up) and can be condensed (down). In the condensed table the empty boxes have disappeared. The value of $S_{CHSH}$ has not changed.

In a table as Fig.3 (down), it is possible to demonstrate that $S_{CHSH} \leq 2$ for all possible values of $a_i, a'_i, b_i, b'_i \neq 0$ and for series of any length [6]. I briefly review the demonstration below:

$$\left| \Sigma a_i . b_i - \Sigma a_i . b'_i \right| = \left| \Sigma a_i . (b_i - b'_i) \right| \leq \Sigma |a_i| . |b_i - b'_i| = \Sigma |b_i - b'_i| \quad (2)$$

where the sum is over the condensed table's length ($N$=4 in Fig.3). In the same way:

$$\left| \Sigma a'_i . b_i + \Sigma a'_i . b'_i \right| \leq \Sigma |b_i + b'_i| \quad (3)$$

summing up eqs.2 and 3:

$$\left| \Sigma a_i . b_i - \Sigma a_i . b'_i \right| + \left| \Sigma a'_i . b_i + \Sigma a'_i . b'_i \right| \leq \Sigma |b_i - b'_i| + \Sigma |b_i + b'_i| \quad (4)$$

For a given value of $i$, the first term in the rhs is 2 (0) if $b_i$ and $b'_i$ have the same (different) sign. The opposite occurs for the second term in the rhs, so that the rhs is 2 for all values of $i$. Hence:

$$(1/N) . \left| \Sigma a_i . b_i - \Sigma a_i . b'_i \right| + (1/N) . \left| \Sigma a'_i . b_i + \Sigma a'_i . b'_i \right| \leq 2 \quad (5)$$

Compare this derivation of CHSH with the usual one, which involves the definition of classical probabilities (which presupposes a Boolean algebra and sufficiently long series) for hypothetical counterfactually definite hidden variables, the statistical independence of these probabilities between stations A and B, and the existence of Riemann or Lebesgue integrals over the space of the hidden variables. These are elaborated, and hence vulnerable and arguable assumptions. Here, instead, all that is assumed is Sica's condition applied to recorded series of outcomes of any length. These series are certainly *real*. They are bits in a classical computer memory, or even signs in a piece of paper. In turn, Sica's condition plays the role of assuming separable phenomena [10]. I find Sica's approach refreshing. It sweeps away the mist produced by years of discussions about the definitions of "Locality" and "Realism".

In summary:

*Sica's condition valid* $\Rightarrow$ *series can be reordered* $\Rightarrow$ *tables can be condensed* $\Rightarrow$ $S_{CHSH} \leq 2$ (6)

The Clauser-Horne inequality (CH) [9] is demonstrated in a similar way. In this case there is only one detector per station, say the "+" ones (see Fig.1). The relevant terms in the inequality are:

$$J \equiv N_c(\alpha,\beta) + N_c(\alpha,\beta') + N_c(\alpha',\beta) - N_c(\alpha',\beta') - S(\alpha) - S(\beta) \quad (7)$$

where $N_c(i,j)$ is the number of "+,+" coincidences when the angle setting is $\{i,j\}$ and $S(\alpha)$ ($S(\beta)$) is the number of single detections in station A(B) with A=$\alpha$ (B=$\beta$). Then $a_i, b_i = \{0,1\}$ only, and $S(\alpha) = \Sigma a_i$ ($S(\beta) = \Sigma b_i$) where the sum is over the total number of time slots in the run. Using Sica's condition, then:

$$J = \Sigma (a_i.b_i + a_i.b'_i + a'_i.b_i - a'_i.b'_i - a_i - b_i) \equiv \Sigma T_i \quad (8)$$

For an arbitrary term $T_i$ in the sum: $T_i = a_i.(b_i + b'_i) + a'_i.(b_i - b'_i) - a_i - b_i$. If $a_i = 0$ and $a'_i = 1$, then $T_i = -b'_i \leq 0$ (recall $b'_i = 0$ or 1 now). If $a_i = a'_i = 0$, then $T_i = -b_i \leq 0$. If $a_i = 1$ and $a'_i = 0$, then $T_i = b'_i -1 \leq 0$. Finally, if $a'_i = a_i = 1$ then $T_i = b_i -1 \leq 0$. Therefore, $T_i \leq 0 \ \forall i \Rightarrow J \leq 0$, and the CH inequality is demonstrated.

## 2. Violation of Bell's inequalities.

The point is that reordering the series is not always possible. This occurs, precisely, when $S_{CHSH} > 2$. This result is derived immediately from inverting the implications in eq.6 but, in order to see how it works, let consider the simplest case, see Figure 4.

Let suppose that $a_i=(-,+)$ is measured for time slots $i=1,2$ (i.e., when B=$\beta'$), and that $b'_i=(+,-)$ is simultaneously measured, so that E($\alpha,\beta'$)=-1. In order to hold to Sica's condition, $a_i$ for $i=3,4$ (i.e., when B=$\beta$) must be also (-,+). This does not necessarily occur in reality, but it can be achieved by reordering the elements of the actually measured series $a_i$. Let suppose then that in $i=3,4$ the series $b_i$ is simultaneously measured to be (-,+), so that E($\alpha,\beta$)=1. In order to hold to Sica's condition, $b_i$ measured in $i=5,6$ must be reordered (if necessary) to be (-,+) too. Let suppose that $a'_i$ is simultaneously measured to be (-,+) in $i$=5,6 so that E($\alpha'$,$\beta$)=1. At $i$=7,8, $a'_i$ is reordered (if necessary) to (-,+) to fit Sica's condition. At this point, *if $b'_i$ is simultaneously measured to be (-,+)*, then E($\alpha'$,$\beta'$)=1 and $S_{CHSH}$=4. But, in this case, it would be $b'_i$=(+,-) for time slots $i$=1,2 (A=$\alpha$) and $b'_i$=(-,+) for time slots $i$=7,8 (A=$\alpha'$), violating Sica's condition. There is no possibility of further reordering.

| Set\ $i$ | 1 | 2 | 3 | 4 | 5 | 6 | 7 | 8 |
|---|---|---|---|---|---|---|---|---|
| $a_i$ | - | + | - | + | | | | |
| $b_i$ | | | - | + | - | + | | |
| $a'_i$ | | | | | - | + | - | + |
| $b'_i$ | +<br>(-) | -<br>(+) | | | | | - | + |

Figure 4: Possible table of factual series in black (distribution of time as in the upper part of Fig.2) with $S_{CHSH}$ = 4. Sica's condition does not hold (see the series $b'_i$ at $i$=1,2 and $i$=7,8). Results in brackets (and red color) are obtained after *arbitrary* reordering the series, i.e., ignoring the *restriction*.

If Sica's condition is enforced to hold, by *arbitrarily* (i.e., ignoring the restriction) imposing $b'_i$ to be (-,+) in time slots $i$=1,2 (in brackets, in red in Fig.4) without simultaneously reordering $a_i$, then E($\alpha$,$\beta'$)=1 and $S_{CHSH}$=2. Instead, if $a_i$ is simultaneously reordered to hold to the restriction, then all series must be reordered in cascade to hold to Sica's condition, and at the end it is reached $b'_i$=(+,-) for time slots $i$=7,8. As $b'_i$ =(-,+) was forced for slots $i$=1,2, Sica's condition is violated again (and $S_{CHSH}$ = 4).

The conclusion is that the original series in Fig.4 ($S_{CHSH}$=4) cannot be made to hold to Sica's condition after legitimate reordering (i.e., reordering obeying the restriction). In this sense, these series are "not classical". Nevertheless, the difference between the series of outcomes $b'_i$ when $i$=1,2 (A=$\alpha$) and when $i$=7,8 (A=$\alpha'$) does not need a "spooky action at a distance" to be explained. The difference is explained simply because the series are measured at different times. There is no intuitively compelling reason for Sica's condition to hold for series recorded at different times, and hence, to conclude that it must be $S_{CHSH}$ ≤ 2. It might be said that *time* plays here the role of the context [9].

In summary: in Sica's approach, the experimentally observed violation of Bell's inequalities is explained by giving up the assumption that (legitimate) reordering the series to fit Sica's condition is always possible. In my opinion, giving up this assumption is more acceptable to intuition than giving up Locality or Realism. It just means that series recorded at different times can be irreducible to each other.

**3. In what sense quantum non-locality does exist.**

If $S_{CHSH}>2$ is observed then, by inverting the relationships in eq.6, Sica's condition does not hold. Violation of Sica's condition means that the series in one station is different if the setting in the other station is changed: the series $a_i$ of outcomes in A changes if the setting β in B is changed (and vice versa). This has indeed the flavor of a non-local effect. In my opinion, the difference between the series, depending on the setting in the other station, is the clearest and most undisputable way to show that "quantum non-locality" does somehow exist (against my previous beliefs) when Bell's inequalities are violated.

Yet, be aware that the difference is between factual and counter-factual series, that is: the series $a_i$ that is in fact observed when B=β, is different from the counter-factual series "$a_i$" that would have been observed if it had been B=β' instead (during the *same* time period). This difference is fatally unobservable, because doing two different, distinguishable actions at the same time (setting B=β and B=β') is a logical impossibility. Time does not go back. This is, also in my opinion, the clearest way to understand why "quantum non-locality" cannot be used to send signals.

Note that this reasoning does not involve hidden variables, probabilities, statistical independence, counterfactual definiteness, quantum algebra, no-cloning theorem, no-signaling boxes or any of the elaborated (and hence, vulnerable) arguments involved in the usual discussions about quantum non-locality. The conclusions derived from Sica's approach are, in consequence, more reliable than the ones derived from the usual discussions on the subject.

**4. Filling the empty boxes.**

In Fig.2, the sums in eqs.2 and 3 that are used to demonstrate the CHSH inequality are disjoint. Therefore, *nothing* can be said about the value of $S_{CHSH}$ [11]. No bound can be established, other than the tautology $S_{CHSH} \leq 4$. In order to put the elements of the series under the same sum, so as to derive some non trivial bound, the empty boxes (which correspond to counter-factual values) must be somehow filled. In turn, assigning numerical values to these empty boxes requires the definition of a "possible world" to ensure logical consistency [10]. There is no mystery in this. Defining a possible world just means that information, additional to what is actually observed, is somehow provided. In the usual derivation using probabilities, the Bell's inequalities are retrieved, or not, depending on the possible world chosen [12]. In Sica's approach instead, filling the empty boxes with *any* outcomes belonging to the set {+,-} leads to a table as the one in the lower part of Fig.3, for which the Bell's inequalities are demonstrated to be valid. If the empty boxes are filled with "0" instead (which implies efficiency η=½) then the inequality becomes $S_{CHSH} \leq 4$, which is *not* violated by Quantum Mechanics' predictions (in general, $S_{CHSH} \times \eta \leq 2$ [13]).

Let consider now the following "possible world": in Figure 5, the factual series (in black, no brackets) $b'_i$ are (+,-) when A=α (*i*=1,2), and (-,+) when with A=α' (*i*=7,8), i.e., Sica's condition is violated (and $S_{CHSH} = 4$). But it is possible to assign counter-factual values (between brackets and in blue color) $b'_{3,4}$=(-,+) and $b'_{5,6}$=(+,-) so that the completed series (which now includes both factual and counter-factual values) is $b'_i$=(+,-,-,+) for both A=α (*i* from 1 to 4) and A=α' (*i* from 5 to 8), then holding to Sica's condition. In the same way, the other completed series in Fig.5 are $b_i$=(-,+,-,+) (*i* from 1 to 4 and from 5 to 8), $a_i$ = (-,+,-,+) (for both B=β, i.e. *i* from 3 to 6, and B=β', i.e. *i* = 1,2 and 7,8) and $a'_i$ = (+,-,-,+). This table is made of both factual and counter-factual outcomes and it does hold to Sica's condition. It is named here a *complete Sica's table*. In the Appendix it is demonstrated that, given any set of factual series, that may violate CHSH or not (as in Fig.4, black or red), it is always possible to draft a complete Sica's table (as Fig.5), where CHSH is (obviously) not violated. The complete Sica's table plays then the role of a classical set of instructions, or classical hidden variable. The violation of CHSH by the factual series can be interpreted as the consequence of having chosen convenient observation times in some classical set of instructions.

| Set\ *i* | 1 | 2 | 3 | 4 | 5 | 6 | 7 | 8 |
|---|---|---|---|---|---|---|---|---|
| $a_i$ | - | + | - | + | (-) | (+) | (-) | (+) |
| $b_i$ | (-) | (+) | - | + | - | + | (-) | (+) |
| $a'_i$ | (+) | (-) | (+) | (-) | - | + | - | + |
| $b'_i$ | + | - | (-) | (+) | (+) | (-) | - | + |

Figure 5: Possible *complete Sica's table* drafted from Fig.4. Factual values are in black and are the same as in Fig.4, they lead to $S_{CHSH}$=4. Counter-factual values (between brackets, in blue) are chosen so that Sica's condition for the complete series (factual + counterfactual) is valid. F.ex.: $b'_i$ = (+,-,-,+) when A=α, *i* = 1 to 4, and also when A=α', *i* = 5 to 8. This table can be now condensed (see Fig.6).

No complete Sica's table can explain the violation of CHSH for all choices of observation times. F.ex., suppose that the factual observation of the setting {α,β'} is chosen to be at time slots *i*=3,4 instead of *i*=1,2 in Fig.5. Then $a_i$ = (-,+) and $b'_i$ = (-,+), E(α,β')=1 (instead of -1) and CHSH is not violated. A successful complete Sica's table can be built only *after* the factual series (that violate CHSH) are defined. Therefore, a complete Sica's table provides a classical *explanation* of the violation of CHSH in an experimental run, but it cannot provide a satisfactory (i.e., violating CHSH) classical *prediction* of factual and counter-factual outcomes for all arbitrary, still unperformed experiments. This impossibility of presenting a classical table of data able to reproduce the predictions of Quantum Mechanics for all possible choices of unperformed experiments is similar to the Kochen-Specker theorem. There is a difference, though: here the impossibility is related with choices of time assignments of observations, instead of choices of outcomes of observations.

As said, a table that holds to Sica's condition has redundant information and can be condensed. Therefore, a complete Sica's table can be condensed in spite of being drafted from a factual series that does violate CHSH. Yet, the condensed table obtained in this way is necessarily a mixture of both factual and counter-factual outcomes. See, f.ex., Figure 6.

| $Set \backslash i$ | 1 | 2 | 3 | 4 |
|---|---|---|---|---|
| $a_i$ | - | + | - | + |
| $b_i$ | (-) | (+) | - | + |
| $a'_i$ | (+) | (-) | - | + |
| $b'_i$ | + | - | (-) | (+) |

Figure 6: Condensed table from Fig.5. Note it includes both factual and counter-factual outcomes.

**Summary.**

The usual derivation of Bell's inequalities using hidden variables and classical probabilities is replaced by considering time series of outcomes. This approach makes easily visible the (often disregarded) unavoidable limitation that measuring with different settings requires measuring at different times. The condition that the series recorded in one station is the same regardless the setting in the other station (Sica's condition) allows the derivation of Bell's inequalities. This can be done in all cases only if *arbitrary* reordering the series (i.e., reordering without restriction) is supposed possible. Giving up this supposition is (in my opinion) more acceptable to intuition than the usual alternative of giving up Locality or Realism. It means accepting that series recorded at different times can be irreducibly different, a sort of "time contextuality". It is worth mentioning that a supposition equivalent to "time *non*-contextuality" has been long ago recognized to be tacitly present, in addition to Locality and Realism, in the usual derivation of Bell's inequalities. It has received different names: ergodicity [14], homogeneous dynamics [11], uniform complexity [15], counter-factual stability [16], and probably more.

If Sica's condition holds in a given table of actually measured outcomes, then the table can be reordered and condensed, and Bell's inequalities are not violated. Inversely, if Bell's inequalities are observed to be violated, then the table of outcomes cannot be condensed, the series cannot be reordered, and Sica's condition does not hold. That is, the series in A (B) are irreducibly different if the setting in B (A) changes. This inescapable result gives the words "quantum non-locality" a precise meaning. Besides, it explains why this "quantum non-locality" cannot be used for faster than light signaling. It cannot, because the difference mentioned above is between factual and counter-factual series, and observing counter-factual series is a logical impossibility.

In the Appendix is demonstrated that, given any factual series of outcomes (as in Fig.4) it is possible to choose counter-factual outcomes such that the complete table holds to Sica's condition (as in Fig.5). There are $2^{N/2}$ of these tables for each set of factual series, where $N/2$ is the length of the series actually recorded for each local setting in each station (f.ex. in Fig.2, $N/2 = 4$). Therefore, even the series that do violate CHSH can be interpreted as the consequence of having chosen appropriate observation times from a classical table (= a table that holds to Sica's condition). As a complete Sica's table always exists (actually, a lot of them), it is possible to conclude that the factual (observed) world of the experiment in Fig.1 may be quantum or not (i.e., Bell's inequalities may be violated or not), but that the union of factual and counter-factual worlds is always classical (i.e., it always holds to Bell's inequalities).

In summary, Sica's approach demonstrates in a simple and hardly disputable way that quantum non-locality does exist (in the factual + counterfactual world which, besides, is classical), the impossibility of using it for signaling (because the counterfactual world is inaccessible), and that series of events occurring within "time" can be intrinsically unrepeatable (otherwise, Bell's inequalities could never be violated).

**Appendix.**

*Proposition: A complete Sica's table can be draft from any set of factual series of outcomes.*

The demonstration is constructive.

Any distribution of the time intervals assigned to the settings (say, a random one, as in the loophole-free experiments) can be reordered (without violating the restriction) to the distribution in the upper part of Fig.2. Therefore, the discussion can deal with this particular distribution of time intervals. This choice has no other purpose and effect than to make notation simpler. For that distribution then, the set of factual (actually observed) series of outcomes is, for each angle setting in each station:

$$(a_1...a_{N/2}); (a'_{N/2}...a'_N); (b_{N/4}...b_{3N/4}); (b'_1...b'_{N/4}) \cup (b'_{3N/4}...b'_N)$$

where, for simplicity, all series are assumed of equal length ($N/2$). The counter-factual series (indicated in bold typing and blue color) are (see upper part of Fig.2):

$$(\mathbf{a_{N/2}...a_N}); (\mathbf{a'_1...a'_{N/2}}); (\mathbf{b_1...b_{N/4}}) \cup (\mathbf{b_{3N/4}...b_N}); (\mathbf{b'_{N/4}...b'_{3N/4}})$$

The complete series (factual ∪ counter-factual) in Alice when A=α and B=β is:

$$(a_{N/4}...a_{N/2}) \cup (\mathbf{a_{N/2}...a_{3N/4}}) \tag{A1}$$

and when B=β' is:

$$(a_1...a_{N/4}) \cup (a_{3N/4}...a_N) \qquad (A2)$$

In the same way, the complete series when A=α' and B=β is:

$$(a'_{N/4}...a'_{N/2}) \cup (a'_{N/2}...a'_{3N/4}) \qquad (A3)$$

and when B=β':

$$(a'_1...a'_{N/4}) \cup (a'_{3N/4}...a'_N) \qquad (A4)$$

When B=β and A=α:

$$(b_1...b_{N/4}) \cup (b_{N/4}...b_{N/2}) \qquad (A5)$$

when B=β and A=α':

$$(b_{N/2}...b_{3N/4}) \cup (b_{3N/4}...b_N) \qquad (A6)$$

when B=β' and A=α:

$$(b'_1...b'_{N/4}) \cup (b'_{N/4}...b'_{N/2}) \qquad (A7)$$

and finally, when B=β' and A=α':

$$(b'_{N/2}...b'_{3N/4}) \cup (b'_{3N/4}...b'_N) \qquad (A8)$$

Sica's condition requires that A1=A2, A3=A4, A5=A6 and A7=A8.

In order to make A1=A2, $(a_1...a_{N/4})$ is reordered to be equal to $(a_{N/4}...a_{N/2})$. To obey the restriction, $(b'_1...b'_{N/4})$ is also reordered and becomes: $(b'_1...b'_{N/4})^R$. The counter-factual (empty boxes) series $(a_{N/2}...a_{3N/4})$ and $(a_{3N/4}...a_N)$ remain free, but must be chosen to be equal between them in order to make A1=A2. F.ex.: one series is freely chosen (there are hence $2^{N/4}$ possible choices for arbitrarily balanced series), and the other one is made equal to the freely chosen one.

In order to make A3=A4, $(a'_{N/2}...a'_{3N/4})$ is reordered to be equal to $(a'_{3N/4}...a'_N)$, and hence the series $(b_{N/2}...b_{3N/4})$ changes to $(b_{N/2}...b_{3N/4})^R$. As before, the series $(a'_{N/4}...a'_{N/2})$ and $(a'_1...a'_{N/4})$ must be equal between them but are otherwise free, so there are again $2^{N/4}$ possible choices.

In order to make A5=A6, it suffices to choose the counter-factual series $(b_1...b_{N/4})$ (which are free) equal to the factual series $(b_{N/2}...b_{3N/4})^R$, and $(b_{3N/4}...b_N)$ equal to $(b_{N/4}...b_{N/2})$. In the same way, to make A7=A8, $(b'_{N/4}...b'_{N/2})$ is chosen equal to $(b'_{3N/4}...b'_N)$ and $(b'_{N/2}...b'_{3N/4})$ equal to $(b'_1...b'_{N/4})^R$. A complete Sica's table is therefore built, and the proposition is demonstrated.

There are $2^{N/4} \times 2^{N/4} = 2^{N/2}$ complete Sica's tables for each set of factual series, where $N/2$ is the length of the series actually recorded for each local setting in each station. F.ex, in Fig.2, $N/2=4$.

**Acknowledgements.**

This work received support from grants PIP 00484-22 and PUE 229-2018-0100018CO, both from CONICET (Argentina).

**References.**

[1] A.Garuccio and F.Selleri, "Nonlocal interactions and Bell's inequality", *Nuovo Cim.* **36B** p.176 (1976).
[2] M.Kupczynski, "Closing the Door on Quantum Nonlocality", *Entropy* **2018**, 20, 877.
[3] A.Khrennikov, "Get Rid of Nonlocality from Quantum Physics", *Entropy* **2019**, 21, 806.
[4] H.Zwirn, "Non Locality vs. modified Realism" *Found.Phys.* **50** p.1 (2020).
[5] A.Hnilo, "Non-Boolean Hidden Variables model reproduces Quantum Mechanics' predictions for Bell's experiment", *arXiv/quant-ph/:2005.10367*
[6] L.Sica, "Bell's inequalities I: an explanation for their experimental violation", *Opt. Commun.* **170** p.55 (1999), and "Bell's inequalities II: Logical loophole in their interpretation", *ibid* p.61.
[7] G.Weihs *et al.*, "Violation of Bell's inequality under strict Einstein locality conditions", *Phys. Rev. Lett.* **81** p.5039 (1998).
[8] M.Agüero *et al.*, "Time stamping in EPRB experiments: application on the test of non-ergodic theories"; *Eur.Phys.J. D* **55** p.705 (2009).
[9] J.Clauser and A.Shimony, "Bell's theorem: experimental tests and implications", *Rep. Prog. Phys.* **41** p.1881 (1978).
[10] B.d'Espagnat, "Nonseparability and the tentative descriptions of reality", *Phys. Rep.* **110** p.201 (1984).
[11] A.Hnilo, "Using measured values in Bell's inequalities entails at least one hypothesis additional to Local Realism", *Entropy* **2017**, 19, 180.
[12] A.Hnilo, "Time weakens the Bell's inequalities"; *arXiv:1306.1383v2*.
[13] A.Hnilo, "Some consequences of Sica's approach to Bell's inequalities", *arXiv/* 2403.03236.
[14] V.Buonomano, "A limitation on Bell's inequality", *Annales del'I.H.P. Sect.A*, **29** p.379 (1978). one hypothesis additional to Local Realism", *Entropy* **19**,80 doi: 103390/e19040180 (2017).
[15] Hou Shun Poh *et al.*, "Probing the quantum–classical boundary with compression software", *New J. Phys.* **18** 035011 (2016).
[16] A.Khrennikov, "Buonomano against Bell: nonergodicity or nonlocality?", *Int. J. Quantum Inf.* **15** 1740010 (2017).